\documentclass[lettersize,journal]{IEEEtran}
\usepackage{amsmath,amssymb,amsfonts}
\usepackage{algorithmic}
\usepackage{graphicx}
\usepackage{textcomp}
\usepackage{xcolor}
\usepackage{subfigure}

\usepackage{bm}
\usepackage{algorithm}
\usepackage{algorithmic}

\usepackage[compact]{titlesec}
\titlespacing{\section}{2.0pt}{*2.0}{*0}
\titlespacing{\subsection}{1.1pt}{*1.1}{*0}
\titlespacing{\subsubsection}{0.3pt}{*0}{*0}

\def\BibTeX{{\rm B\kern-.05em{\sc i\kern-.025em b}\kern-.08em
    T\kern-.1667em\lower.7ex\hbox{E}\kern-.125emX}}
\usepackage{soul}
\usepackage{balance}
\usepackage{cite}
\usepackage[version=3]{acro}  
\DeclareAcronym{ICI}{short = ICI ,long  = inter-carrier interference ,tag = abbrev}
\DeclareAcronym{OFDM}{short = OFDM ,long  = orthogonal frequency division multiplexing ,tag = abbrev}
\DeclareAcronym{DFT}{short = DFT ,long  = discrete Fourier transform ,tag = abbrev}
\DeclareAcronym{IDFT}{short = IDFT ,long  = inverse \ac{DFT} ,tag = abbrev}
\DeclareAcronym{6G}{short = 6G ,long  = sixth generation ,tag = abbrev}
\DeclareAcronym{5G}{short = 5G ,long  = fifth generation ,tag = abbrev}
\DeclareAcronym{4G}{short = 4G ,long  = fourth generation ,tag = abbrev}
\DeclareAcronym{DD}{short = DD ,long  = delay-doppler ,tag = abbrev}
\DeclareAcronym{TF}{short = TF ,long  = time-frequency ,tag = abbrev}
\DeclareAcronym{MIMO}{short = MIMO ,long  = multiple input multiple output ,tag = abbrev}
\DeclareAcronym{3GPP}{short = 3GPP ,long  = 3rd generation partnership program ,tag = abbrev}
\DeclareAcronym{MA}{short = MA ,long  = multiple access ,tag = abbrev}
\DeclareAcronym{MUI}{short = MUI ,long  = multi-user interference ,tag = abbrev}
\DeclareAcronym{SINR}{short = SINR ,long  = signal to interference plus noise ratio ,tag = abbrev}
\DeclareAcronym{TDMA}{short = TDMA ,long  = time-division multiple access ,tag = abbrev}
\DeclareAcronym{FDMA}{short = FDMA ,long  = frequency-division multiple access ,tag = abbrev}
\DeclareAcronym{SNR}{short = SNR ,long  = signal to noise ratio ,tag = abbrev}
\DeclareAcronym{FCP}{short = FCP ,long  = full-CP,tag = abbrev}
\DeclareAcronym{HM}{short = HM ,long  = high mobility,tag = abbrev}
\DeclareAcronym{LM}{short = LM ,long  = low mobility,tag = abbrev}
\DeclareAcronym{AWGN}{short = AWGN ,long  =  additive white gaussian noise,tag = abbrev}
\DeclareAcronym{BS}{short = BS ,long  =  base station,tag = abbrev}
\DeclareAcronym{DL}{short = DL ,long  =  downlink ,tag = abbrev}
\DeclareAcronym{UL}{short = UL ,long  =  uplink,tag = abbrev}
\DeclareAcronym{BER}{short = BER ,long  =  bit-error rate,tag = abbrev}
\DeclareAcronym{SC}{short = SC ,long  =  sub-carrier,tag = abbrev}
\DeclareAcronym{BW}{short = BW ,long  =  band-width,tag = abbrev}
\DeclareAcronym{MSE}{short = MSE ,long  =  mean-square-error,tag = abbrev}
\DeclareAcronym{MU}{short = MU ,long  =  multi-user,tag = abbrev}

\DeclareAcronym{ST}{short = ST ,long  =  symbol time,tag = abbrev}
\DeclareAcronym{SCS}{short = SCS ,long  =  sub-carrier spacing,tag = abbrev}
\DeclareAcronym{PEP}{short = PEP ,long  =  pairwise-error probabilities,tag = abbrev}
\DeclareAcronym{UE}{short = UE ,long  =  user equipment,tag = abbrev}
\DeclareAcronym{vic-UE}{short = vic-UE ,long  =  victim UE,tag = abbrev}
\DeclareAcronym{ZP}{short = ZP ,long  =  zero-padding,tag = abbrev}
\DeclareAcronym{CP}{short = CP ,long  =  cyclic prefix,tag = abbrev}
\DeclareAcronym{TMA}{short = TMA ,long  =  timing misalignment,tag = abbrev}
\DeclareAcronym{SBFD}{short = SBFD ,long  =  subband full-duplex,tag = abbrev}
\DeclareAcronym{CLI}{short = CLI ,long  =  cross-link interference,tag = abbrev}
\DeclareAcronym{ISI}{short = ISI ,long  =  inter-symbol interference,tag = abbrev}
\DeclareAcronym{vic}{short = vic ,long  =  victim,tag = abbrev}
\DeclareAcronym{SIR}{short = SIR ,long  =  signal-to-interference ratio,tag = abbrev}
\DeclareAcronym{agg}{short = agg ,long  =  aggressor,tag = abbrev}
\DeclareAcronym{FFT}{short = FFT ,long  =  fast Fourier transform,tag = abbrev}
\DeclareAcronym{NMSE}{short = NMSE ,long  =  normalized mean square error,tag = abbrev}
\DeclareAcronym{IFFT}{short = IFFT ,long  = inverse fast Fourier transform,tag = abbrev}
\DeclareAcronym{TDD}{short = TDD ,long  = time division duplexing,tag = abbrev}
\DeclareAcronym{TA}{short = TA ,long  = timing advance,tag = abbrev}

\DeclareAcronym{LP}{short = LP ,long  = low-power,tag = abbrev}

\DeclareAcronym{OOK}{short = OOK ,long  = on-off keying,tag = abbrev}

\DeclareAcronym{PHY}{short = PHY ,long  =physical layer,tag = abbrev}

\DeclareAcronym{LS}{short = LS ,long  =least-squares,tag = abbrev}
\DeclareAcronym{LN}{short = LN ,long  =least-norm,tag = abbrev}

\DeclareAcronym{GD}{short = GD ,long  = gradient descent,tag = abbrev}

\DeclareAcronym{SGD}{short = SGD ,long  = stochastic GD,tag = abbrev}

\DeclareAcronym{IoT}{short = IoT ,long  = Internet of Things,tag = abbrev}

\DeclareAcronym{A-IoT}{short = A-IoT,long  = ambient \ac{IoT},tag = abbrev}

\DeclareAcronym{WuR}{short = WuR ,long  = wake-up radio ,tag = abbrev}
\DeclareAcronym{WuS}{short = WuS ,long  = wake-up signal ,tag = abbrev}

\DeclareAcronym{SE}{short = SE ,long  = spectral efficiency,tag = abbrev}

\DeclareAcronym{MDR}{short = MDR ,long  = mis-detection rate,tag = abbrev}

\DeclareAcronym{FAR}{short = FAR ,long  = false alarm rate,tag = abbrev}

\DeclareAcronym{QAM}{short = QAM ,long  = quadrature amplitude modulation,tag = abbrev}

\DeclareAcronym{FD}{short = FD ,long  = frequency-domain,tag = abbrev}

\DeclareAcronym{TD}{short = TD ,long  =  time-domain,tag = abbrev}

\begin{document}

\title{CP-Aware OFDM-based OOK Signaling}

 \author{ Badr Eddine Ouakouak, Salah Eddine Zegrar,~\IEEEmembership{Member,~IEEE}, and H\"{u}seyin Arslan,~\IEEEmembership{Fellow,~IEEE}
\thanks{Badr Eddine Ouakouak, Salah Eddine Zegrar and  H\"{u}seyin Arslan are with the Department of Electrical and Electronics Engineering, Istanbul Medipol University, Istanbul, 34810, Turkey (e-mail: badr.ouakouak@std.medipol.edu.tr; salah.zegrar@medipol.edu.tr; huseyinarslan@medipol.edu.tr).}}

\maketitle

\begin{abstract}
This letter addresses the challenge of cyclic prefix (CP) problem in orthogonal frequency division multiplexing (OFDM)-based on-off keying (OOK) generation for low-power Internet of Things (IoT) systems. We propose a CP-aware waveform design that generates the OOK signal over the entire CP-OFDM symbol duration, potentially breaking subcarrier orthogonality after passing through the channel if the CP is not matched. To enable such a mechanism, OOK has to be generated independently and a modification to the existing OFDM-based OOK generation block is needed. Additionally, a block-wise time shift can be applied to minimize CP mismatch. Analytical discussions and simulations confirm that the resulting interference on legacy subcarriers remains controlled, while OOK detection performance significantly improves, albeit with trade-offs that should be carefully considered in practical deployments.
\end{abstract}

\begin{IEEEkeywords}
OOK, low-power, OFDM, IoT, CP handling.
\end{IEEEkeywords}

\section{Introduction}

The proliferation of \ac{IoT} has led to a growing demand for efficient and reliable communication protocols tailored to extremely \ac{LP} devices. Among various modulation schemes, \ac{OOK} is particularly well-suited for such applications due to its simplicity and minimal power requirements. In this context, integrating \ac{OOK} with \ac{OFDM} offers a promising approach to improve spectral efficiency while preserving the low complexity essential for \ac{IoT} deployments~\cite{3gpp_38_869,3gpp_38_769}.

Recent wireless standards have introduced mechanisms such as \ac{WuS} to enable \ac{LP} operation by allowing devices to deactivate their primary radios during idle periods. These mechanisms can significantly reduce energy consumption, particularly when the \ac{WuS} is implemented using simple modulation schemes like \ac{OOK}~\cite{3gpp_38_869}. On the transmitter side, generating such signals using \ac{OFDM} offers practical advantages, including coexistence with conventional data transmissions and compatibility with existing infrastructure.
These developments are increasingly relevant in the broader context of \ac{A-IoT}, where \ac{LP} devices are expected to operate in dense environments and leverage ambient or opportunistic signaling.

Various strategies have been proposed to generate OFDM-compatible \ac{OOK} waveforms for \ac{LP} applications. Early designs, such as IEEE 802.11ah~\cite{zhang2017low}, use power-level encoding per OFDM symbol, while later work explored peak shaping~\cite{caballe2019alternative}, \ac{TD} matching~\cite{sahin2018sequence}, and \ac{LS}-based subcarrier optimization~\cite{mazloum2020interference}, \cite{zhang2023toward}.
\textcolor{black}{Recent 5G releases specify an \ac{LP} \ac{WuS} using \ac{DFT}-spread-\ac{OFDM}–based OOK \cite{3gpp_38_869}. For \ac{A-IoT}, \ac{LS} is also an option and is left as an implementation choice for the manufacturer \cite{3gpp_38_769}.}
However, the mentioned works often overlook a critical issue involving the \ac{CP} insertion between consecutive OFDM symbols carrying OOK data. Existing CP-handling strategies include discarding the CP~\cite{jian2024ambient}. This may require prior knowledge of CP length~\cite{hoglund2025synchronization} or estimating CP boundaries through edge timing~\cite{AMb_General_aspects}. These methods require accurate synchronization, which is impractical for many \ac{LP} or passive devices, and fail under sampling frequency offsets~\cite{AMb_General_aspects}.
Other approaches mitigate the CP issue by inserting guard zeros~\cite{WUS_WUR_design} or enforcing chip continuity~\cite{AMb_General_aspects}. While these help suppress transition edges and improve OOK detection, they reduce spectral efficiency and degrade overall performance at high chip rates. Therefore, more effective solutions are needed to enable sustainable and reliable OOK communication within existing OFDM frameworks.

The present work addresses the CP mismatch directly at the waveform design stage,  eliminating the need for assumptions at the OOK receivers. Our contributions are summarized as follows: 

\begin{enumerate}
    \item We propose a CP-aware OFDM-based OOK generation scheme that designs the OOK signal over the CP-OFDM symbol duration. Orthogonality is intentionally relaxed, and a time shift is used to minimize CP mismatch and leakage.  
    \item We analytically characterize the relation between CP mismatch energy and the resulting interference on data subcarriers, and show that this leakage can be effectively bounded, albeit with trade-offs that we also discuss.
    \item We validate through simulations that the induced leakage remains non-severe under practical scenarios, and demonstrate the robustness of the proposed scheme in improving OOK detection.  
\end{enumerate}


\section{System Model}


Consider a \ac{DL} \ac{OFDM} system with $N$ subcarriers representing the \ac{DFT} size. Let $s^\text{f}_{k}(m)$ represent the transmitted data symbols for the $k$-th OFDM symbol with $k \in \{0,1,\dots,K-1\}$. Then, the \ac{TD} signal $s^\text{t}_{k}(n)$ is given as 
 \begin{equation}
        s^\text{t}_{k}(n) = \frac{1}{\sqrt{N}}\sum_{m=0}^{N-1}s^\text{f}_{k}(m)e^{j\frac{2\pi mn}{N}}.
    \label{equ:IFFT_DSP}
\end{equation}
This can be expressed in matrix form using the \ac{DFT} matrix $\mathbf{F}$ as follows\footnote{We use the bold lowercase $\mathbf{x}$ to denote a column vector of length $N$, with the entries being the samples of the discrete signal $x(n), n=0,1,...,N-1$. $\mathbf{A}^\dagger$ denotes the Hermitian of the matrix $\mathbf{A}$. $[,\cdot,]_B$ denotes modulo-$B$ indexing.}:
 \begin{equation}
        \mathbf{s}^\text{t}_{k}= \mathbf{F}^\dagger\mathbf{s}^\text{f}_{k}.
    \label{equ:IFFT_matrix}
\end{equation}
In this system, $W$ subcarriers are reserved for generating a \ac{TD} \ac{OOK} waveform intended for transmission to the \ac{LP} device, while the remaining $D = N-W$ subcarriers are reserved for data. 
In parallel with conventional OFDM data targeting legacy \acp{UE}, we aim to transmit an OOK sequence $\mathbf{d}_k = [d_k(0), d_k(1), \ldots, d_k(M-1)]^T \in \{0, 1\}^M$, where $M$ is the number of chips per OFDM symbol\footnote{With Manchester encoding, $M$ doubles the bit count. Regardless of the used encoding, the system operates on chips rather than bits.}. 

Let $c = N/M$ denote the chip duration in samples. The desired \ac{TD} \ac{OOK} signal for the $k$-th symbol, $\mathbf{x}_k$, is obtained by repeating each $d_k(l)$ value $c$ times. The transmitted signal \(x_k(n)\) is constructed as $ x_k(n) =
 d_k(\lfloor n/c \rfloor)$, which can be written in vector form as $\mathbf{x}_k = \mathbf{d}_k \otimes \mathbf{1}_{c \times 1}$, where $\otimes$ denotes the Kronecker product and $\mathbf{1}_{c \times 1}$ is a column vector of ones of length $c$. Then,  the $W$ selected subcarriers are used to synthesize an approximate OOK signal $\mathbf{\Tilde{x}}_k$, using \ac{IDFT}.


Let $\mathbf{a}_k$ be the \ac{FD} coefficients that best approximate $\mathbf{x}_k$. Because the OOK signal uses only $W$ subcarriers, it is given by 
\begin{equation}
        \Tilde{\mathbf{x}}_k =  \Tilde{\mathbf{F}}\mathbf{a}_k,
\end{equation}
\textcolor{black}{where $\Tilde{\mathbf{F}} \in \mathbb{C}^{N \times W}$ consists of the $W$ columns of the $N$-point \ac{IDFT} matrix $\mathbf{F}^\dagger$ corresponding to the $W$ lowest-frequency subcarriers \cite{mazloum2020interference}. This includes negative and positive frequencies, as frequency-domain conjugate symmetry is needed to synthesize a real time-domain OOK signal.}
A straightforward way to compute $\mathbf{a}_k$ is via the \ac{LS} method as 
\begin{equation}
        \mathbf{a}_k =  {(\Tilde{\mathbf{F}}^\dagger\Tilde{\mathbf{F}})}^{-1}\Tilde{\mathbf{F}}^\dagger\mathbf{x}_{k},
        \label{eq:LS_solution}
\end{equation}
which yields the best $W$-subcarrier approximation of the original waveform as
\begin{equation}
        \Tilde{\mathbf{x}}_k =  \Tilde{\mathbf{F}}{(\Tilde{\mathbf{F}}^\dagger\Tilde{\mathbf{F}})}^{-1}\Tilde{\mathbf{F}}^\dagger\mathbf{x}_{k}.
\end{equation}
Note that when $W = N$, the reconstruction becomes exact, i.e., $\Tilde{\mathbf{x}}_k = \mathbf{x}_k$. However, in general, $W < N$, and the approximation incurs distortion. 

\begin{figure}[!t]
\centering
\includegraphics[scale = .2]{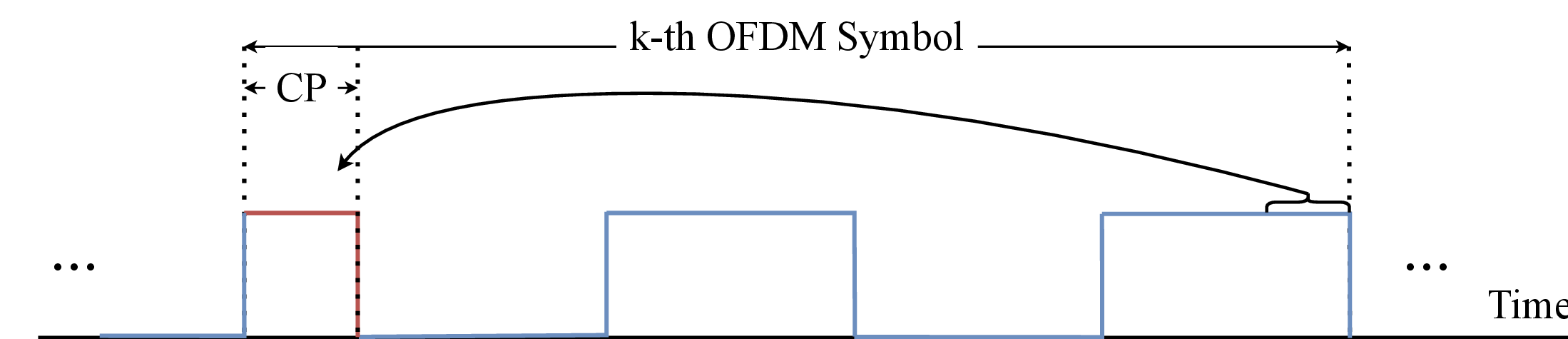}
\caption{Illustration of unintended edge transitions caused by CP insertion.}
\label{fig:issue_caused_by_CP}
\end{figure}

\section{Problem Description and Proposed Method}

\subsection{CP-Handling Problem Description}

In addition to the inherent distortion from subcarrier approximation, a challenge arises due to the insertion of  \ac{CP}. According to standard OFDM operation,  following the synthesis of the approximate \ac{TD} waveform $\Tilde{\mathbf{x}}_k$, a \ac{CP} of length $L$ is prepended to the signal, before transmission. This process may introduce synchronization difficulties, and more critically, it can create spurious edge transitions that distort the intended OOK sequence. As illustrated in Fig.~\ref{fig:issue_caused_by_CP}, CP insertion can produce a false falling edge at the beginning of the symbol, thereby altering the desired bit sequence. Such CP-induced discontinuities constitute a fundamental obstacle to adapting OFDM systems for reliable OOK generation.

\subsection{Proposed CP-Handling Method}

This subsection presents the proposed OFDM-based scheme for OOK signal generation. The design addresses the CP mismatch problem inherent to conventional OOK waveforms, while unavoidably introducing interference to the legacy data subcarriers. As shown in the subsequent evaluations, this interference can be effectively bounded and maintained within controllable limits. 


In this work, we take a block-level view of the OOK waveform. The OOK sequence spans $K$ OFDM symbols and is given by $\mathbf{d} = [d(0), d(1), \ldots, d(KM-1)]^T \in \{0, 1\}^{KM}$. According to the proposed scheme, the OOK time sequence is generated across the duration of each CP-OFDM symbol. Hence, the chip duration becomes $c = (N+L)/M$ samples. Then, the desired \ac{TD} \ac{OOK} signal \(x(n)\) is constructed as
\begin{equation}
    x(n) =
\begin{cases}
    1, & \text{if } d(\lfloor n/c \rfloor) = 1 \\
    0, & \text{otherwise}
\end{cases},~ \text{for}~n \in \{0,1,\dots,B-1\},
\label{equ:me}
\end{equation}
where $B = KN'$ denotes the duration of the OFDM block, and $N' = N+L$ is the duration of one CP-OFDM symbol. Therefore, the transmitted OOK sequence can be expressed as $\mathbf{x} = \mathbf{d} \otimes \mathbf{1}_{c \times 1}$, and  $\mathbf{x} = {[\mathbf{x}^\text{T}_0,...,\mathbf{x}^\text{T}_{K-1}]}^\text{T}$.


\begin{figure}[!t]
  \centering
     \subfigure[ Before shifting.]{\includegraphics[scale=0.15]{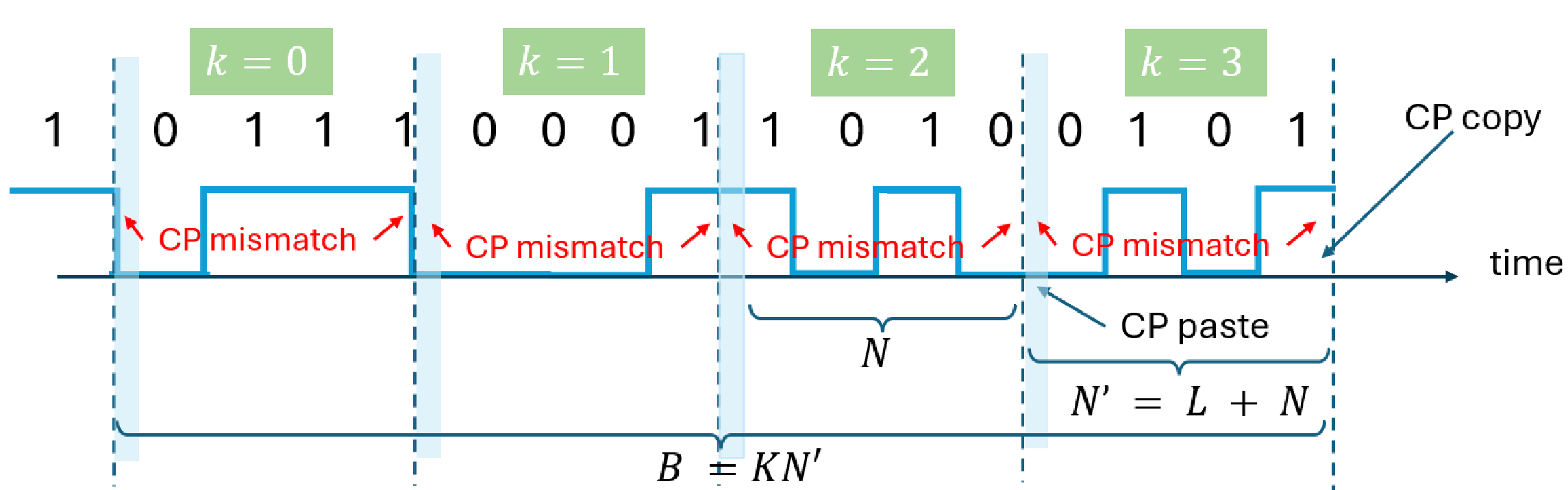}} 
     \subfigure[After shifting with $\sigma_{\text{opt}}$.]{\includegraphics[scale=0.15]{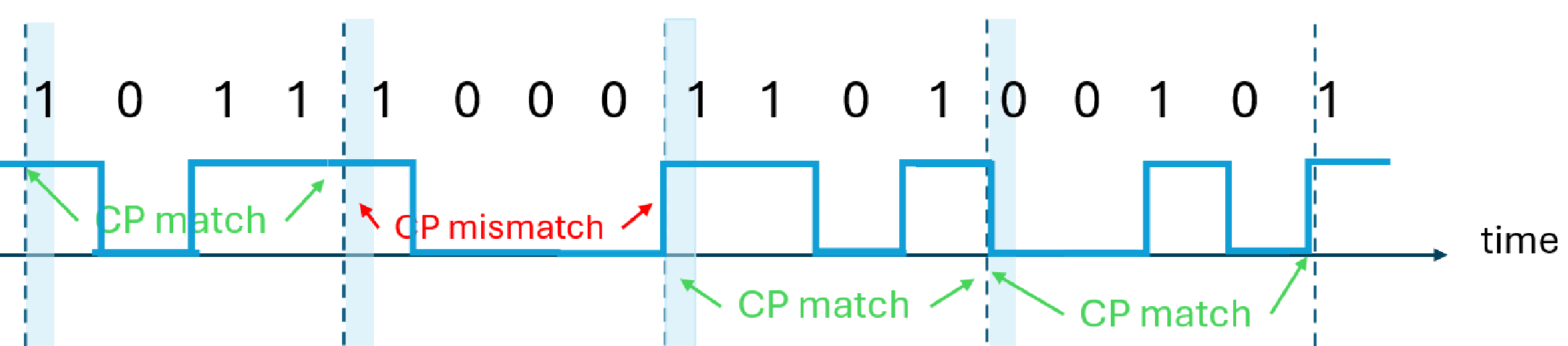}}
  \caption{\textcolor{black}{Minimizing the CP mismatch.}} 
      \label{fig:shiftingproposed}
 \end{figure}

To facilitate the description of the proposed scheme, we introduce the following matrices. Let the CP-addition matrix be $\bm{\Gamma} \triangleq {[[\mathbf{0}_{L\times (N-L)},\mathbf{I}_{L}];\mathbf{I}_{N}]}$, and define the CP-removal matrix $\bar{\bm{\Gamma}}$ as $\bar{\bm{\Gamma}} \triangleq {[\mathbf{0}_{N\times L},\mathbf{I}_{N}]}$. Furthermore, let $\bm{\Gamma}_l$ denote the CP-extracting matrix that selects the first $L$ samples of an $N' \times 1$ vector, and let $\bm{\Gamma}_r$ denote the matrix that extracts the last $L$ samples. These matrices are respectively given by $\bm{\Gamma}_l = {[\mathbf{I}_{L}, \mathbf{0}_{L\times N}]}$, $\bm{\Gamma}_r= {[ \mathbf{0}_{L\times N},\mathbf{I}_{L}]}$.

In order to produce a valid approximation of the desired OOK waveform $\mathbf{x}$, a consistency condition must be satisfied at the CP boundaries. Specifically, the first and last $L$ samples of $\mathbf{x}_k$ must be identical, which we refer to as the \textit{CP condition}: $\bm{\Gamma}_l\mathbf{x}_k = \bm{\Gamma}_r\mathbf{x}_k$. This condition ensures that $\mathbf{x}_k$ can be viewed as a legitimate output of a CP-based OFDM block. However, this constraint is generally not satisfied by arbitrary OOK waveforms. 
As a result, the next objective is to minimize the mismatch between the copied and added CP regions. In conjunction with the earlier block-level formulation, we aim to minimize this discrepancy across all $K$ OFDM symbols in the block. For notational clarity, we refer to $\bm{\Gamma}_r \mathbf{x}_k$ as the \textit{copied CP}, and $\bm{\Gamma}_l \mathbf{x}_k$ as the \textit{added CP}.

The core idea behind maximizing the likelihood of satisfying the CP condition lies in recognizing that the data-bearing OFDM signal and the OOK waveform are inherently independent. This decoupling implies that the OOK signal does not need to be time-aligned with the data \ac{IDFT} timeline.
Leveraging this observation, we introduce a \ac{TD} shift to the OOK signal relative to the OFDM block structure, with the goal of maximizing the similarity between the \textit{copied CP} and the \textit{added CP} across the entire block. Mathematically, we model this shift using the circular shift matrix $\bm{\Pi}_B$ of size B, such that the shifted signal is given by
\begin{equation}
    \mathbf{z}_\sigma = \bm{\Pi}^\sigma_B \mathbf{x},
\end{equation}
where $\sigma$ denotes the shift (in samples)\footnote{We assume block-level repetition is used, which is typical in low-power systems. Thus, inter-block transitions are irrelevant for the current analysis. The method does not inherently rely on this assumption. It is used to make the modeling possible.}.

\begin{figure}[!t]
\centering
\includegraphics[scale = .10]{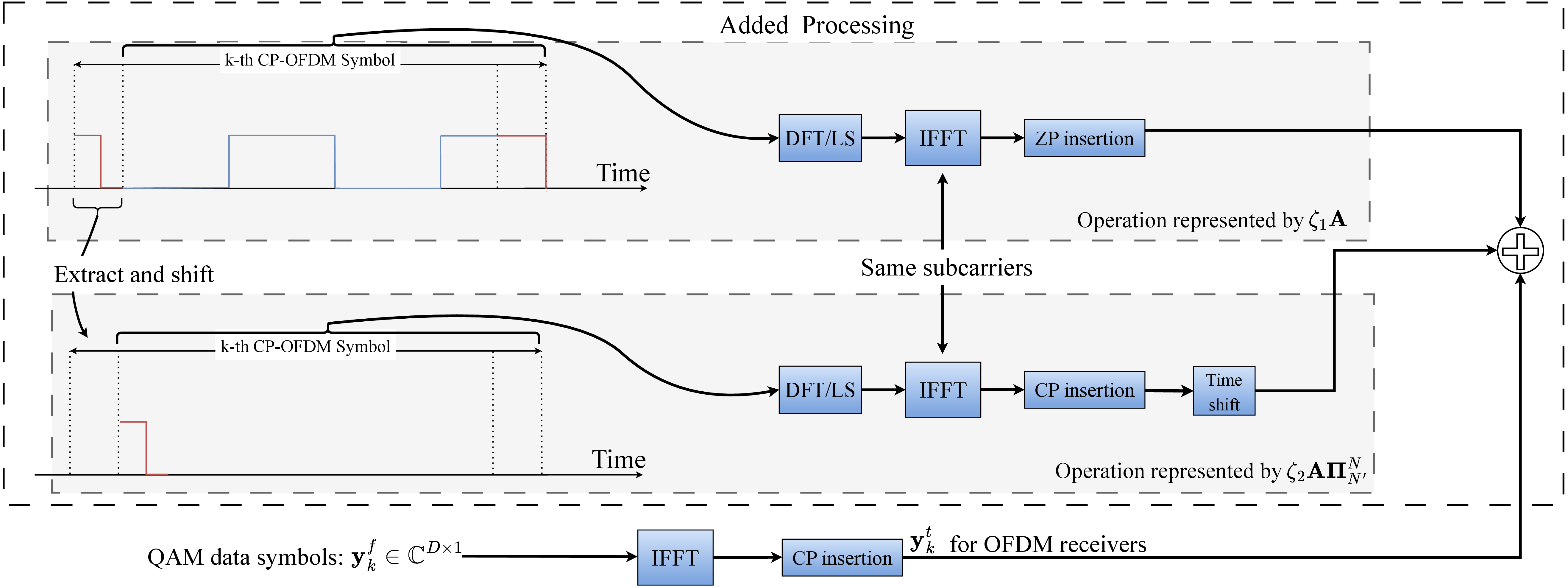}
\caption{\textcolor{black}{OOK Generation according to the proposed scheme.}}
\label{fig:added_complexity_at_the_BS}
\end{figure}

To formalize the idea of minimizing the mismatch between the CP regions, we define the following two vectors: $  \mathbf{v}^\sigma_l = (\mathbf{I}_K \otimes \bm{\Gamma}_l)\, \mathbf{z}_\sigma$, $
    \mathbf{v}^\sigma_r = (\mathbf{I}_K \otimes \bm{\Gamma}_r)\, \mathbf{z}_\sigma
$, which stack, respectively, the \textit{added CP} and \textit{copied CP} samples for all $K$ symbols in the block after the shift.

The optimal value $\sigma_{\text{opt}}$ minimizes the block-wide CP mismatch, which yields an average mismatch energy of
\begin{equation}\small
\begin{aligned}
&\Sigma(\sigma) = \frac{1}{KL}  \| \mathbf{v}^{\sigma}_l - \mathbf{v}^{\sigma}_r \|^2\\
 &= \frac{1}{KL}\sum_{k = 0}^{K - 1}\sum_{l = 0}^{L - 1} \left| z_\sigma(kN'+l) - z_\sigma(kN'+N+l) \right|^2\\
&= \frac{1}{KL}\sum_{k = 0}^{K - 1}\sum_{l = 0}^{L - 1} \left| x([kN'+l-\sigma]_B) - x([kN'+N+l-\sigma]_B) \right|^2.
\end{aligned}
\label{equ:mismatch}
\end{equation}
Then, the optimal shift $\sigma_{\text{opt}}$ is obtained by solving the optimization problem that minimizes $\Sigma$ as follows:
\begin{equation}
\begin{aligned}
\sigma_{\text{opt}} = &\arg\min_{\sigma } \quad   \Sigma(\sigma)\\
     \text{subject to} &\quad  \sigma \in \{0, \dots, c - 1\}.
\end{aligned}
\label{equ:opt2}
\end{equation}
 Once $\sigma_{\text{opt}}$ is determined, the following strategy can be employed for generating the final OOK waveform. \textcolor{black}{Fig.~\ref{fig:shiftingproposed} illustrates the effect of shifting the OOK sequence by $\sigma_{\text{opt}}$ samples in reducing the overall number of CP mismatch instances.}

We proceed with generating the OOK block $\mathbf{x}$ despite knowing that the CP condition is not satisfied. In this case, the CP for the OOK waveform is generated separately and appended to the \ac{TD} signal, which is then superimposed on the CP-OFDM data signal. This separation may disrupt orthogonality, potentially causing interference (or leakage) over the data subcarriers after propagation through a time-dispersive channel. However, the trade-off is shown to offer significant advantages in handling the CP constraint.

The primary source of this leakage lies in the mismatch between the copied and added CP regions. Since the signal $\mathbf{x}$ is synthesized without enforcing the CP condition, these discontinuities cannot be avoided.
To model the resulting OOK waveform, consider the LS-based approximation of the desired block, given by
\begin{equation}
        \Tilde{\mathbf{x}} = (\mathbf{I}_K\otimes\mathbf{A})\mathbf{z}_{\sigma_{\text{opt}}},
        \label{eq:approx_xtild_without_CP}
\end{equation}
where $\mathbf{A} =\tilde{\mathbf{F}}{(\tilde{\mathbf{F}}^\dagger\tilde{\mathbf{F}})}^{-1}\tilde{\mathbf{F}}^\dagger\bar{\bm{\Gamma}}$.

\begin{figure}[!t]
    \centering
    \includegraphics[scale=.35]{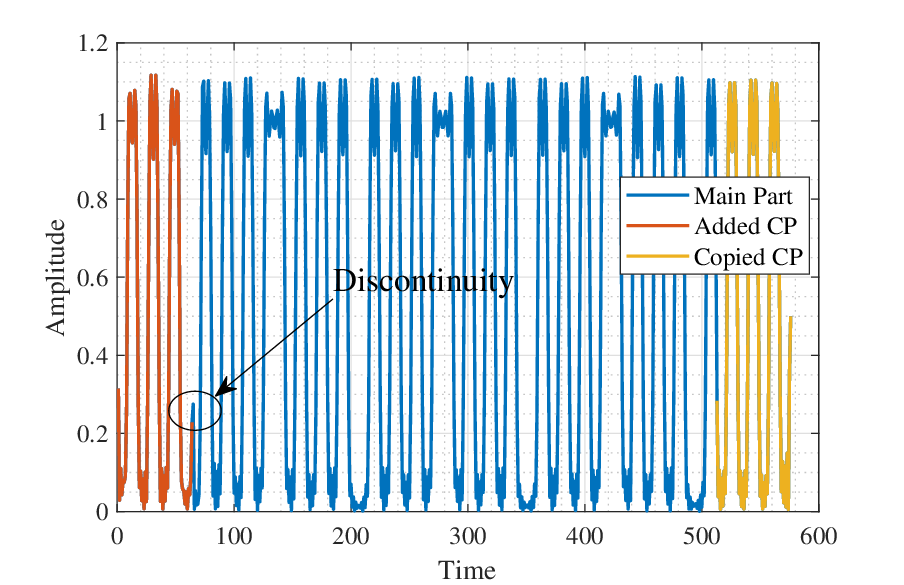}
    \caption{\textcolor{black}{An example illustrating the resulting CP.}}
\label{fig:resultingCP_discontnuity}
\end{figure}

Unlike conventional OFDM, CP addition for the OOK waveform is handled independently from that of the data signal, and thus cannot be expressed as a simple multiplication by $(\mathbf{I}_K \otimes \bm{\Gamma})$. A practical approach is to view CP addition as a transformation applied after a temporal shift. Specifically, the signal is first circularly shifted by $N$ samples, then transformed using \eqref{eq:approx_xtild_without_CP}. To formalize this, define two matrices: $\bm{\zeta}_1$, which appends zeros to the beginning of the OFDM symbol, and $\bm{\zeta}_2$, which inserts the CP and zeros out the main symbol body. These operations are respectively given by $\bm{\zeta}_1 = {[\mathbf{0}_{L \times N};\mathbf{I}_{N}]}$ and $\bm{\zeta}_2 =  {[[\mathbf{0}_{L\times (N-L)},\mathbf{I}_{L}];\mathbf{0}_{N \times N}]}$. It can then be shown that the actual generated OOK signal $\mathbf{u}$ is given by
\begin{equation}
       {\mathbf{u}} =  \bigg(\mathbf{I}_K\otimes(\overbrace{\bm{\zeta}_1\mathbf{A}}^\text{main part}+\underbrace{\bm{\zeta}_2\mathbf{A}\bm{\Pi}^{N}_{N'}}_\text{CP part})\bigg)\mathbf{z}_{\sigma_{\text{opt}}}.
\end{equation}
\textcolor{black}{Fig.~\ref{fig:added_complexity_at_the_BS} summarizes the proposed process of OOK generation. The OOK CP is independently generated and added before the CP-OFDM data signal.} 
Accordingly, the added processing involves two OFDM-based OOK blocks (one for the main part and one for the CP part). The OOK generation involves two $W$-DFT processes and two $N$-IDFT operations, as opposed to only one $W$-DFT processes and one $N$-IDFT operation in the conventional system. While the BS can handle this processing, backward compatibility is still an issue that warrants further investigation. \textcolor{black}{Finally, Fig.~\ref{fig:resultingCP_discontnuity} illustrates the effect of minimizing the mismatch with a practical (approximate) OOK signal. As can be seen, the added CP and copied CP are not perfectly matched, but the mismatch level between them is small. There is also a discontinuity at the beginning of the main part, which is caused by the independent generation of the CP. This is one of the drawbacks of the proposed method, which also needs to be taken into consideration.}

\section{Performance Analysis}

In this section, we relate the size of the mismatch $\Sigma(\sigma)$ to the amount of interference experienced by the data subband due to the CP condition not being satisfied. This is done in order to bound the interference and potentially reduce the
average leakage from the OOK subcarriers over the data subband. First, intuitively, we know that if $\mathbf{v}^\sigma_l=\mathbf{v}^\sigma_r$ there is no leakage. In fact, the part of the added CP $\mathbf{v}^\sigma_l = \mathbf{v}^\sigma_l-\mathbf{v}^\sigma_r +\mathbf{v}^\sigma_r$ that causes the interference is given only by $\mathbf{v}^\sigma_l-\mathbf{v}^\sigma_r $, since $\mathbf{v}^\sigma_r$ causes no interference (i.e., it is the matched part). To give an expression on this interference, define the matrix $\bm{\zeta}$, which appends $N$ zeros at the end of a column vector of size $L$. This matrix is defined as $\bm{\zeta} = {[\mathbf{I}_{L };\mathbf{0}_{N \times L}]}$. The interference is then given by the DFT of the signal $(\mathbf{I}_K\otimes \bm{\zeta})(\mathbf{v}^\sigma_l-\mathbf{v}^\sigma_r)$, after passing through the frequency-selective channel and CP removal.
\begin{figure}[!t]
    \centering
    \includegraphics[scale=.35]{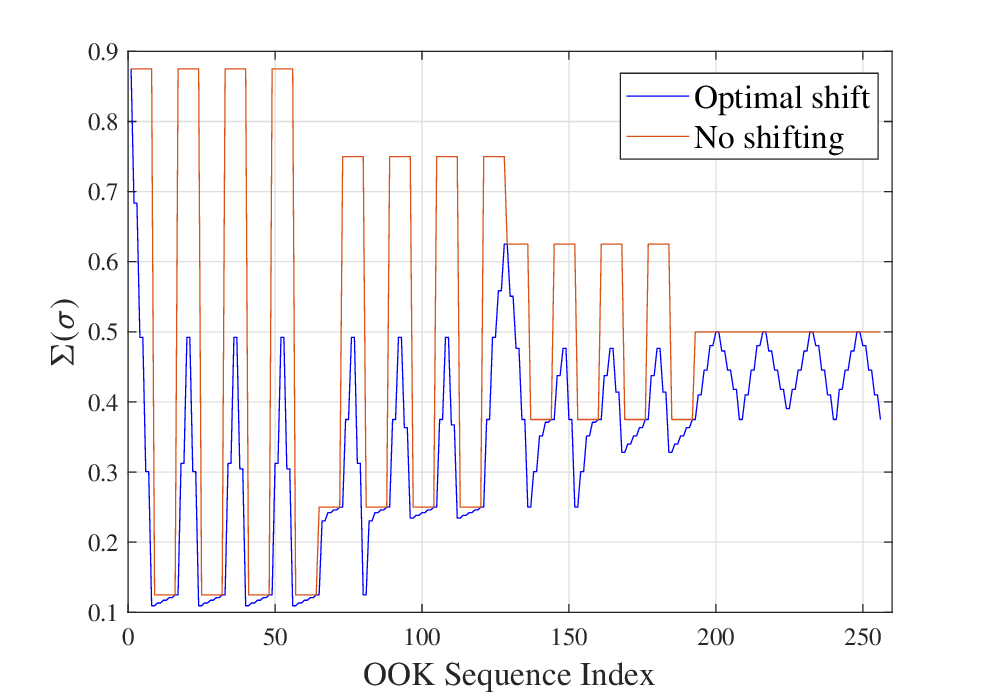}
    \caption{Amount of Mismatch for different OOK sequences.}
\label{fig:amount_of_mismatch_for_different_sequenes}
\end{figure}

The vector $\bm{\phi}$, representing the interference over the entire block,  is given by
\begin{equation}
       {\bm{\phi}} =  \sum_{i=0}^{p-1}h_i(\mathbf{I}_K\otimes \mathbf{F}\bar{\bm{\Gamma}}\bm{\Pi}^{l_i}_{N'}\bm{\zeta})(\mathbf{v}^\sigma_l-\mathbf{v}^\sigma_r),
       \label{equ:dsds}
\end{equation}
where $h_i$ is the channel coefficient associated with the $i$-th path with delay $l_i$, and $p$ denotes the number of paths.
As seen in \eqref{equ:dsds}, the norm of ${\bm{\phi}}$ is only related to the norm of $(\mathbf{v}^\sigma_l-\mathbf{v}^\sigma_r)$, and it can only change size by changing $(\mathbf{v}^\sigma_l-\mathbf{v}^\sigma_r)$. It also depends on the channel delays, but these are out of the transmitter's and receiver's control.
Let $\mathbf{v}_\sigma=(\mathbf{I}_K\otimes\bm{\zeta})(\mathbf{v}^\sigma_l-\mathbf{v}^\sigma_r)$. Using the mixed-product property, $\bm{\phi}$ can be rewritten as 
\begin{equation}
       {\bm{\phi}} =  \sum_{i=0}^{p-1}h_i(\mathbf{I}_K\otimes \mathbf{F}\bar{\bm{\Gamma}}\bm{\Pi}_{N'}^{l_i})\mathbf{v}_\sigma.
       \label{eq:received_signal}
\end{equation}
With that, we can find ${|\bm{\phi}|}^2 =\bm{\phi}^\dagger\bm{\phi} $ as
\begin{equation}
       {\|\bm{\phi}\|}^2 =  \bigg(\sum_{i=0}^{p-1}h_i(\mathbf{I}_K\otimes \mathbf{F}\bar{\bm{\Gamma}}\bm{\Pi}_{N'}^{l_i})\mathbf{v}_\sigma\bigg)^\dagger\sum_{i=0}^{p-1}h_i(\mathbf{I}_K\otimes \mathbf{F}\bar{\bm{\Gamma}}\bm{\Pi}_{N'}^{l_i})\mathbf{v}_\sigma.
\end{equation}
This can be shown to equal
\begin{equation}
\begin{aligned}
     {\|\bm{\phi}\|}^2 &=   \sum_{j=0}^{p-1}\sum_{i=0}^{p-1}h_j{h_i}^*\mathbf{v}^\dagger_\sigma(\mathbf{I}_K\otimes \bm{\Pi}_{N'}^{-l_j}\bar{\bm{\Gamma}}^\dagger\bar{\bm{\Gamma}}\bm{\Pi}_{N'}^{l_i})\mathbf{v}_\sigma\\
     &=\sum_{j=0}^{p-1}\sum_{i=0}^{p-1}h_j{h_i}^* {(\mathbf{I}_K\otimes \bar{\bm{\Gamma}}\bm{\Pi}_{N'}^{l_j}\mathbf{v}_\sigma)}^\dagger(\mathbf{I}_K\otimes \bar{\bm{\Gamma}}\bm{\Pi}_{N'}^{l_i}\mathbf{v}_\sigma). 
\end{aligned}
       \label{eq:interference_exact}
\end{equation}
As can be seen in \eqref{eq:interference_exact}, not only does the leakage from the OOK to the data band depend on the specific sequence of OOK, it also depends on channel's delay values. These are random and generally uncontrollable. However, using Cauchy-Schwartz's inequality, we can give an upper bound on ${\|\bm{\phi}\|}^2$ that shows the significance of $\Sigma(\sigma)$ and which is independent of the channel's specific delays. The interference also depends on the CP size represented in $\bar{\bm{\Gamma}}$, as it helps in further reducing the leakage, by absorbing the channel delays.
 
Since $\bm{\Pi}_{N'}$ is an orthonormal matrix, it preserves the Euclidean norm. By the Cauchy–Schwarz inequality, we have
\begin{equation}
       {\|\bm{\phi}\|}^2 \leq {\|\mathbf{v}_\sigma\|}^2\sum_{j=0}^{p-1}\sum_{i=0}^{p-1}h_j{h_i}^* = {KL\cdot\Sigma(\sigma)}{|\sum_{j=0}^{p-1}\sum_{i=0}^{p-1}h_j{h_i}^*|}. 
\end{equation}
While the interference depends heavily on the values of channel delays, we are able to estimate the upper bound and relate it to $\Sigma(\sigma)$. This confirms the anticipated importance of the latter. By reducing the value of $\Sigma(\sigma)$, we are effectively improving the bound on the interference and potentially reducing the average leakage from the OOK subcarriers over the data sub-band. 
Let $\phi_{\text{up}} = KL\cdot\Sigma(\sigma){|\sum_{j=0}^{p-1}\sum_{i=0}^{p-1}h_j{h_i}^*|}$. Since $\Sigma(\sigma)$ is minimized if and only if $\sigma = \sigma_{\text{opt}}$, the interference is also minimized when $\sigma = \sigma_{\text{opt}}$.
Finally, the average upper bound per OFDM symbol $\phi_{p}$ is given by ${\phi_{\text{up}}}/{K}$, which spreads over the data subcarriers in the corresponding OFDM symbol.

\section{Simulation Results}
For the numerical evaluations\footnote{While it is not required that the OFDM symbol contain an integer number of chip durations, since that greatly limits the data rate values, the resulting OOK frame should contain an integer number of chips. According to the proposed scheme $K(N+L)/c$ should be an integer (as it indicates the number of chips transmitted in one frame). For the conventional scheme $KN/c$ should be an integer.}, we adopt a CP length of  $L = \frac{N}{8}$. Let $M = 32$, $K = 8$, $N = 512$, and $W = 64$. We consider a setup based on 256 arbitrary OOK sequences.  

Fig.~\ref{fig:amount_of_mismatch_for_different_sequenes} illustrates the amount of mismatch ($\Sigma(\sigma)$) before and after applying the optimal shift for each of the OOK sequences under consideration. As can be seen, the mismatch can be significantly improved after shifting, for most of the sequences. In addition, no shifting is needed for some sequences. For others, the shift does not reduce the mismatch significantly. Due to the strong dependency of mismatch on the specific bit pattern, we select three representative sequences from the previously considered set. These include a worst-match sequence (index 1), referred to as Seq1; a moderate-to-poor match sequence (index 185), referred to as Seq2; a good-to-moderate-match sequence (index 112), referred to as Seq3.

\begin{figure}[t]
  \centering
     \subfigure[ SINR performance.]{\includegraphics[scale=0.375]{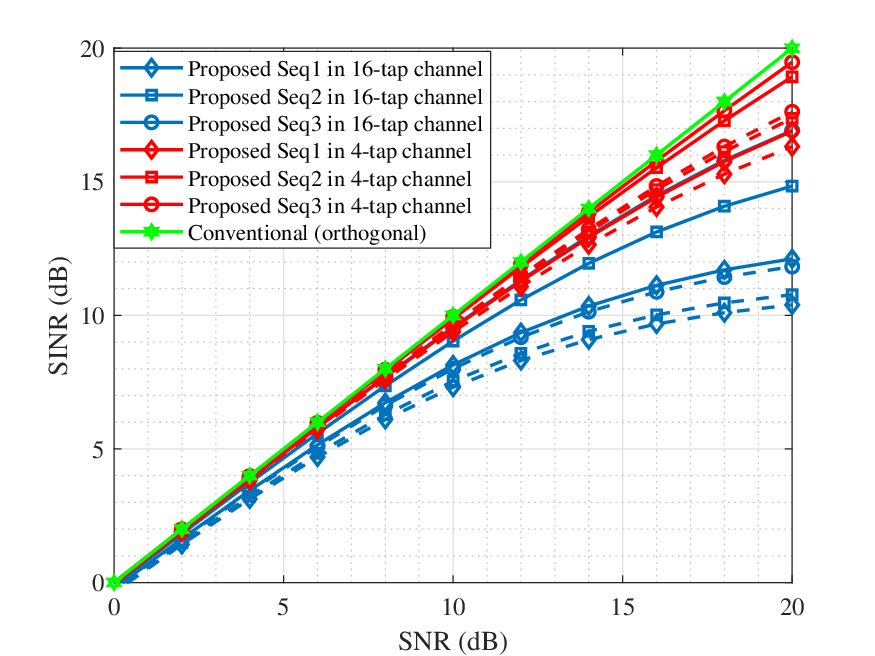}} 
     \subfigure[BER performance.]{\includegraphics[scale=0.375]{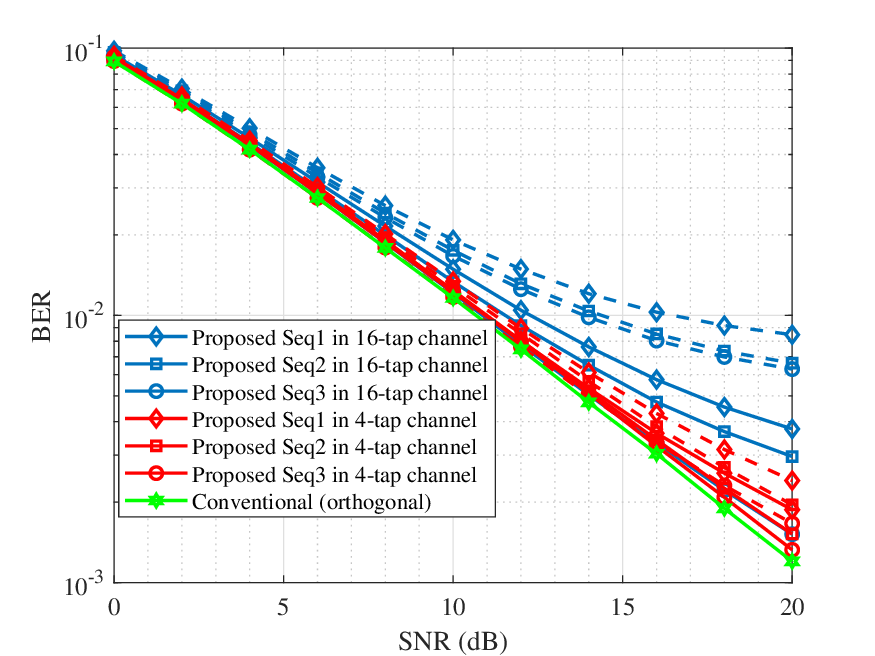}}
  \caption{\textcolor{black}{Performance of the OFDM data receivers.}} 
      \label{fig:SINR_performance}
 \end{figure}
 
The impact of orthogonality loss in the proposed system on legacy data users is evaluated in Fig.~\ref{fig:SINR_performance} \textcolor{black}{(solid lines)} in terms of the \ac{SINR} and \ac{BER} of OFDM users in the presence of OOK transmission. The OOK generation is done according to the proposed scheme (in red and in blue), benchmarked against the conventional orthogonal  scheme (in green). As can be seen in Fig.~\ref{fig:SINR_performance}(a), the SINR is more degraded for larger number of channel taps. For 4-taps, the leakage is quite tolerable.  However, we are interested in the worst-case SINR drop (Seq1), which is about 10 dB at SNR = 20 dB for 16-tap channel. This indicates the limitations of the proposed scheme. The worst-case leakage is much less severe at moderate SNR, not exceeding 1 dB at SNR = 4 dB.
It is also interesting to see the effect on the BER performance for the OFDM users, Fig.~\ref{fig:SINR_performance}(b). Here, the worst-case performance loss is around $0.5\%$ at an SNR of 20 dB, and it is less evident at lower SNR values. \textcolor{black}{Before turning to OOK receiver performance, we also investigate the impact of perfectly rectangular OOK signals (as defined in \eqref{equ:me}) on the resulting interference from CP. We show this performance in dashed lines in Fig.~\ref{fig:SINR_performance}. As expected, the sharp transitions in the ideal OOK signal make the interference effect more pronounced compared to the realistic smoother OOK signal. This means that the proposed method fits better with practical OOK generation than with an ideal rectangular OOK signal.}

\begin{figure}[!t]
    \centering
    \includegraphics[scale=.35]{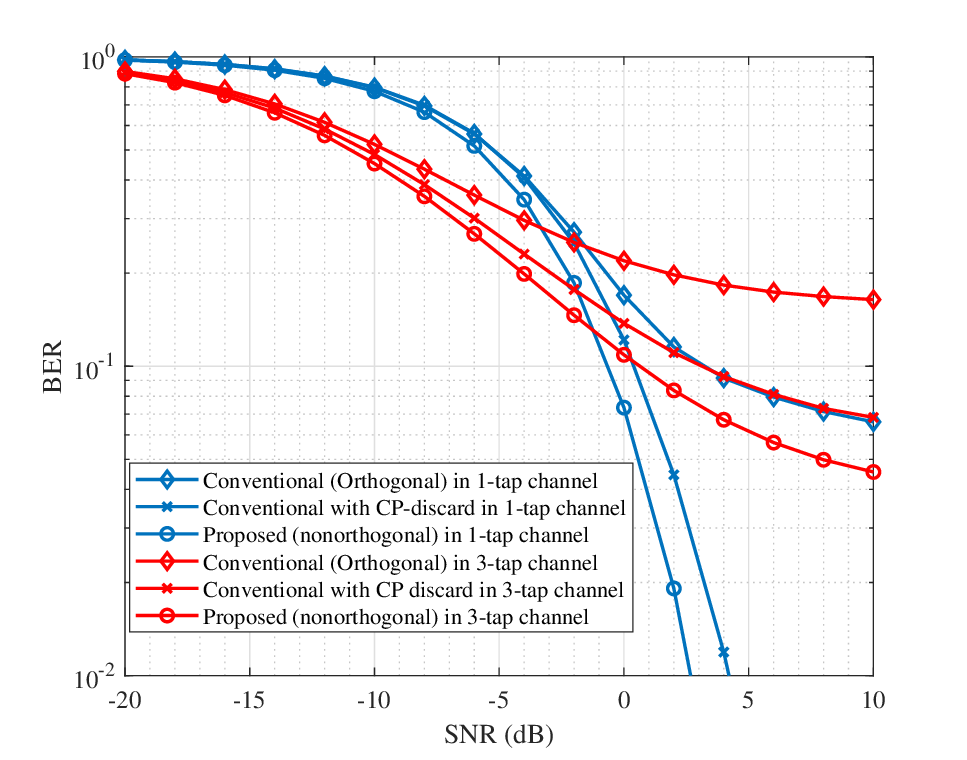}
    \caption{BER performance of the OOK receivers.}
\label{fig:BER_performance}
\end{figure}

The performance of the OOK receiver under the CP mismatch problem, which constitutes the primary motivation for the proposed method, is evaluated in  Fig.~\ref{fig:BER_performance} for both single-tap channel and 3-tap channel. The gain is clearly seen in the proposed scheme, which gives the same BER for all OOK sequences, as it does not rely on the likelihood of CP match. For all schemes, we are showing the average performance across all sequences. The conventional method of OOK generation in an orthogonal manner (with no CP removal at the receiver) shows a clear error floor both for single-tap and multi-tap channel. This is because any false edge transition is interpreted as erroneous bit, thereby setting a bound on the number of correct bits that can be decoded in the duration of an OFDM symbol. In real implementation, CP removal at the receiver can be done by estimating the edge transition intervals, and it depends on the device's sampling frequency offset. We have included the performance of such a system. Clearly it significantly enhances the BER, and no error floor is present for single-tap channel. However, the proposed system still benefits from longer chip durations, which results in a gain in the BER compared to this scheme.

\section{Conclusion}

This letter presented a CP-aware OFDM-based OOK generation scheme that eliminates CP distortion by extending the OOK waveform over the entire CP-OFDM symbol duration. While the method relaxes subcarrier orthogonality and requires block-level modifications, analytical and simulation results show that the resulting interference remains controlled and that OOK detection performance is notably improved. The approach offers a promising direction for low-power IoT systems, though its practical adoption must account for the inherent trade-offs highlighted in this study.
\section*{Acknowledgment}
This study was supported in part by The Scientific and Technological Research Council of Türkiye (TUBITAK) under the Grant Number 124N803. The authors thank to TUBITAK for their supports.

\bibliographystyle{IEEEtran}
\bibliography{IEEEfull}

@article{mazloum2020interference,
  title={{Interference-free OFDM embedding of wake-up signals for low-power wake-up receivers}},
  author={Mazloum, Nafiseh and Edfors, Ove},
  journal={IEEE Trans. Green Commun. Netw.},
  volume={4},
  number={3},
  pages={669--677},
  year={2020},
  publisher={IEEE}
}

@inproceedings{sahin2018sequence,
  title={{Sequence-based OOK for orthogonal multiplexing of wake-up radio signals and OFDM waveforms}},
  author={Sahin, Alphan and Yang, Rui},
  booktitle={2018 IEEE GLOBECOM},
  pages={1--6},
  year={2018},
  organization={IEEE}
}

@techreport{3gpp_38_869,
  title = {{Study on low-power wake-up signal and receiver for NR (Release 18)}},
  institution = {3GPP},
  type = "{Technical Report}",
  number = "TR 38.869",
  year = "2023",
  month = "December",
}

@techreport{3gpp_38_769,
  title = {{Study on solutions for ambient IoT (Internet of Things) (Release 18)}},
  institution = {3GPP},
  type = "{Technical Report}",
  number = "TR 38.769",
  year = "2024",
  month = "November",
}

@article{zhang2023toward,
  title={{Toward zero-energy devices: Waveform design for low-power receivers}},
  author={Zhang, Ticao and others},
  journal={IEEE Commun. Lett.},
  volume={27},
  number={8},
  pages={2038--2042},
  year={2023},
  publisher={IEEE}
}

@article{zhang2017low,
  title={{A low-power OFDM-based wake-up mechanism for IoE applications}},
  author={Zhang, Hualei and others},
  journal={IEEE Trans. Circuits Syst. II: Express Briefs},
  volume={65},
  number={2},
  pages={181--185},
  year={2017},
  publisher={IEEE}
}

@article{caballe2019alternative,
  title={{An alternative to IEEE 802.11 ba: Wake-up radio with legacy IEEE 802.11 transmitters}},
  author={Caball{\'e}, Mart{\'\i} Cervi{\`a} and others},
  journal={IEEE Access},
  volume={7},
  pages={48068--48086},
  year={2019},
  publisher={IEEE}
}

@inproceedings{jian2024ambient,
  title={{Ambient IoT: Insight and Challenge of Enabling Technologies for Future Study}},
  author={Jian, Rongling and others},
  booktitle={2024 ECNCT},
  pages={451--458},
  year={2024},
  organization={IEEE}
}

@misc{hoglund2025synchronization,
  title={{Synchronization for a Communication Node}},
  author={H{\"o}glund, Andreas and others},
  year={2025},
  month=jan # "~2",
  publisher={Google Patents},
  note={{US} Patent App. 18/712,637}
}

@misc{AMb_General_aspects,
  title        = "{{Ambient IoT General aspects of physical layer design}}",
  howpublished = "3GPP TDoc R1-2409311, Meeting",
  month        = oct,
  year         = 2024,
  note         = "{\url{https://portal.3gpp.org/}, accessed April 2025}"
}

@misc{WUS_WUR_design,
  title        = "{{LP-WUS and LP-SS design}}",
  howpublished = "3GPP TDoc R1-2408537, Meeting",
  month        = oct,
  year         = 2024,
  note         = "{\url{https://portal.3gpp.org/}, accessed April 2025}"
}
\vfill
\end{document}